\documentclass[twocolumn, usenatbib]{aastex701}
\usepackage{graphicx}	
\usepackage{amsmath}	
\usepackage{amssymb}
\usepackage{comment}
\usepackage{color}

\newcommand{\rem}[1]{ } 

\begin{document}

\title{Synchrotron-cooled plasma distribution in the outer magnetosphere of a neutron star}
\author[0000-0001-5987-2856]{Mikhail V. Medvedev}
\email[show]{medvedev@ku.edu}
\affiliation{Department of Physics and Astronomy, University of Kansas, Lawrence, KS 66045}
\affiliation{Laboratory for Nuclear Science, Massachusetts Institute of Technology, Cambridge, MA 02139}
\author[0000-0001-9179-9054]{Anatoly Spitkovsky}
\email{anatoly@astro.princeton.edu}
\affiliation{Department of Astrophysical Sciences, Princeton University, Princeton, NJ 08544}
\author[0000-0001-7801-0362]{Alexander Philippov}
\email{sashaph@umd.edu}
\affiliation{Department of Physics, University of Maryland, College Park, MD 20742}
\affiliation{Institute for Research in Electronics and Applied Physics, University of Maryland, College Park, MD 20742}

  

\begin{abstract}
The guiding center formalism is employed to analyze the motion of a charged relativistic particle in an inhomogeneous magnetic field, subject to magnetic mirroring and energy loss due to cooling. The governing equation for the evolution of the magnetic moment is derived. An example representing a neutron star (pulsar or magnetar) magnetosphere is presented to illustrate typical particle orbits. Notably, radiative losses are most pronounced near a trapped particle's turning point. Depending on the initial particle's pitch angle, energy loss can become catastrophic, resulting in the rapid migration of the particle into the loss cone and subsequent precipitation onto a neutron star. Conversely, particles with a larger pitch angle remain temporarily trapped and form a gradually decaying ``cooled-loss-cone" or ``funnel'' distribution, characterized by the maximum momentum space particle density being located at the edge of the loss cone. The size of the loss cone is energy-dependent and scales as $\alpha_{c} \propto \gamma^{3/10}$. Synchrotron losses are strongest in a well-localized region of the magnetosphere, about a few hundred to a thousand star radii under typical pulsar and magnetar conditions. This region is a plausible site for synchrotron radiation originating in the outer magnetosphere, and could also be responsible for non-polar coherent pulsar emission, as well as weak fast radio bursts.
\end{abstract}

\keywords{plasmas -- relativistic processes -- magnetic trapping -- mirroring -- radiative cooling}


\section{Introduction}
\label{s:intro}

Recent advances in numerical modeling of pulsars and magnetars \citep{Spitkovsky06, Tchekhovskoy+13, Philippov+14, Philippov+15, Cerutti+16,  CB17,Philippov+18, Kalapotharakos+18, Chen+20, Hakobyan+23, Parfrey+24, Koushik+26} call for better theoretical understanding of a joint action of elementary processes such as magnetic mirroring and radiative cooling \citep{Bilbao+23, Bilbao+24, Lyutikov+25}. The theory and observations of pulsars and magnetars are nicely summarized in reviews by \citet{PhK22} and \citet{KB17}.

Neutron star magnetospheres are typically described by a dipolar field. This field can naturally confine energetic charged particles within a closed zone due to the magnetic mirroring effect toward polar regions. Such a particle population is observed and known as the ``van Allen belts" in the case of the Earth's magnetosphere. Neutron star magnetospheres should be no exception, albeit with a much stronger magnetic field. This implies that the trapped population undergoes rapid and substantial energy loss (cooling). However, the distribution function of the trapped population and its temporal evolution remain areas of limited scientific understanding. 

Furthermore, plasma can be produced in the vicinity of the magnetospheric current sheet and then propagate along open magnetic field lines toward the neutron star, moving into regions where the magnetic field strength increases. Magnetic mirroring can dynamically influence such plasmas, their evolution, and distribution function. Radiative cooling, when sufficiently rapid, can further modify the plasma dynamics of its inward flow.  

Radiative cooling of the magnetospheric plasma is accompanied by the emission of synchrotron radiation, which, as we will demonstrate, occurs far from the star's surface and can potentially be observed as non-polar emission. There is observational evidence of radiation coming from outer magnetospheres of millisecond and young pulsars \citep{PhK22, KJ26}. Similarly, radiative cooling, as we will illustrate, results in a particle distribution function characterized by loss cones, where the phase space particle density is diminished drastically. Such anisotropic distributions are generally unstable with respect to various instabilities. Notably, they are capable of generating maser emission \citep{GZh70, Zh97book}. It is plausible that such a maser mechanism can produce coherent synchrotron emission, which can be observed as weak fast radio bursts, such as FRB~200428 associated with the galactic magnetar, SGR~1935+2154, \citep{magnetar-frb}.

The main technical concept of our approach is to ``integrate out" the rapid and uninteresting gyro-motion of a particle while preserving the impact of cyclotron or synchrotron cooling \citep{MM21, MM22, MM23a, MM23b}. In essence, our approach employs the guiding center formalism, which allows us to treat a particle as a ``Larmor particle" located at the center of its Larmor orbit, possessing the same electric charge and a specific magnetic moment. In the conventional guiding center formalism, the magnetic moment is conserved, serving as an adiabatic invariant. However, once cooling is factored in, the magnetic moment ceases to be constant. In this paper, we derive the equation governing the evolution of the magnetic moment due to radiative energy loss. Notably, our approach is applicable to any loss mechanism that does not generate a recoil force parallel to the magnetic field, as measured in the frame co-moving with the Larmor particle. We further compute the pertinent particle distributions, their evolution, and their structure.

The paper is structured as follows. In Section \ref{gc}, we provide a concise overview of the classical guiding center equations. Subsequently, in Section \ref{mu-rl}, we derive the evolution equation for the magnetic moment in a strong field, considering the effects of energy loss. In Section \ref{ensemble}, we delve into the evolution of a distribution function for an ensemble of particles. Finally, in Section \ref{bottle}, we illustrate the particle evolution on an example of a straight magnetic bottle, which serves as a simplified model of a dipolar magnetosphere. Here we also investigate the structure and dynamics of the particle distribution function and present interesting analytical estimates. Concluding remarks are presented in Section \ref{concl}.

\section{Relativistic particle motion in a converging magnetic field}
\label{gc}

Particle motion in a magnetic field is characterized by the fast gyro-motion in the plane perpendicular to the magnetic field and the motions associated with its parallel propagation and perpendicular drifts. In neutron star and magnetar magnetospheres, the spatial scale of the field inhomogeneity, which is greater than the neutron star (NS) size, $L\gtrsim R_{NS}$, is many orders of magnitude larger than the Larmor scale, $r_{L}$, associated with gyro-motion. In such a case, the small Larmor scale can be integrated out, which often simplifies further analysis. Upon averaging particle's equations of motion over the fast gyro-motion, one obtains the set of equations describing  the motion of the center of gyration (i.e., the `guiding center'). The approximation when the fast Larmor motion of a particle is averaged out is called the ``drift approximation.'' For a relativistic particle moving in an inhomogeneous field, the guiding center equations are given by \citet{Vandervoort60, N63, Ripperda+18} and described in detail by \citet{Sivukhin65}.

We are interested in the motion of relativistic particles in the neutron star magnetosphere. Suppose a group of particles is injected near the equator and starts to move toward one of the magnetic poles. For a given field line, the field at the poles is stronger than at equator. Hence these polar magnetic mirrors will reflect particles back to the equator. The reflection point of a particle depends on its energy and initial pitch-angle (the angle between the particle's momentum and the local direction of the magnetic field). 

The curvature of the field line would induce both the gradient and curvature drifts in the direction perpendicular to the local gradient of the magnetic field strength and to the local field curvature direction, respectively.  However, the velocities of these drifts are very small, $v_{d}\sim c(r_{L}/L)\gamma_{\perp,\|}\ll c$, and can be safely ignored in our future analysis. Indeed, we can assume that the components of particle's velocity are $v_{\perp}\sim v_{\|}\sim c$, so the nonrelativistic electron Larmor radius is $r_{L}=mc^{2}/eB\sim10^{3}B^{-1}$~cm (where $B$ is in gauss), $L=B/|\nabla B|$ is the characteristic size (e.g., radius) of the system, and $\gamma_{\|},\ \gamma_{\perp}$ are the Lorentz factors associated with the motion parallel to the magnetic field and in the plane perpendicular to it. For concreteness, we estimate that in megagauss fields, which are of interest to us (see below), $r_{L}\sim 10^{-3}$~cm, whereas the corresponding size is $L\sim 10^{8}$~cm for a typical pulsar. Consequently, $v_{d}\ll c$ is a safe assumption for leptons with $\gamma\ll 10^{11}$. For simplicity, it is convenient to assume that the magnetic field is purely converging, i.e., locally radial and straight. The guiding center equations now read:
\begin{eqnarray}
\frac{d\bar{\bf r}}{dt}&=&v_{\|} \hat{\bf b}, 
\label{R}\\
\frac{d\left(m\gamma v_{\|}\right)}{dt}&=&-\frac{\mu_{\rm r}}{\gamma} \hat{\bf b}\cdot\nabla B,
\label{v}\\
\frac{d\left(m\gamma^{2} v_{\bot}^{2}/2B\right)}{dt}&=&\frac{d\mu_{\rm r}}{dt}=0,
\label{mu}
\end{eqnarray}
where $\bar{\bf r}$ is the position of the guiding center, $v_{\|}={\bf v}\cdot  \hat{\bf b}$ and $v_{\bot}$ are the parallel and perpendicular components of the particle velocity (note, the parallel velocity of the guiding center is equal to the particle's ${v_{\|}}$), $B=\left|{\bf B}\right|$ is the magnetic field strength, $\hat{\bf b}={\bf B}/B$ is the unit vector along the magnetic field,  $m$ is the particle mass and $\gamma=1/\sqrt{1-v^{2}/c^{2}}$ is its Lorentz factor with $v^{2}=v_{\|}^{2}+v_{\bot}^{2}$. Note that the relativistic magnetic moment 
\begin{equation}
\mu_{\rm r}=\frac{m\gamma^{2} v_{\bot}^{2}}{2B}
\label{mur}
\end{equation}
differs from its non-relativistic definition $\mu=K_{\bot}/B$ with $K_{\bot}$ being the perpendicular kinetic energy and reduces to it when $\gamma\to1$. Also note that $\nabla B\|\hat{\bf b}$ in the assumed converging field configuration. Equations (\ref{R})--(\ref{mu}) describe the evolution of the guiding center position $R$, the relativistic parallel momentum $m\gamma v_{\|}$, and the relativistic magnetic moment $\mu_{\rm r}$, respectively.

\section{Magnetic moment with radiative losses}
\label{mu-rl}

Deep inside the neutron star magnetosphere, the magnetic field is strong, so that synchrotron losses can no longer be neglected. Synchrotron cooling reduces the perpendicular energy of a relativistic particle. The magnetic moment is no longer constant and  decreases with time, so Eq.~(\ref{mu}) should be modified. The total energy of the particle and radiation is constant, $E_{\rm tot}=\gamma mc^{2}+E_{\rm rad}=const$. Differentiation with respect to time yields, $dE_{\rm tot}/dt=d\left(\gamma mc^{2}\right)/dt+P=0$, where $P=dE_{\rm rad}/dt$ is the radiation emission power. In the lab frame, in which Eqs.~(\ref{R})--(\ref{mu}) are written, the energy equation becomes 
\begin{equation}
\frac{d\left(\gamma mc^{2}\right)}{dt}=-P.
\label{E}
\end{equation}
For synchrotron emission 
\begin{equation}
P=P_{\rm sync}=\frac{2}{3}\frac{r_{e}^{2}}{c}v_{\bot}^{2}\gamma^{2}B^{2}
=\frac{\sigma_{T}}{4\pi c}v_{\bot}^{2}\gamma^{2}B^{2},
\label{P}
\end{equation}
where $r_{e}\equiv e^{2}/mc^{2}$ is the classical electron radius and $\sigma_{T}\equiv(8\pi/3)r_{e}^{2}\simeq 6.65\times10^{-25}\textrm{ cm}^{2}$ is the Thomson cross-section.

Equation (\ref{E}) can be re-written as
\begin{equation}
mc^{2}\left(\frac{\partial\gamma}{\partial v_{\|}^{2}}\frac{d v_{\|}^{2}}{d t}
+\frac{\partial\gamma}{\partial v_{\bot}^{2}}\frac{d v_{\bot}^{2}}{d t}\right)=-P.
\end{equation}

We note that only perpendicular motion is affected by radiative (synchrotron or cyclotron) energy loss, hence ${d v_{\|}^{2}}/{d t}=0$. Indeed, in the comoving reference frame ($v_{\|}=0$), the radiation emission is symmetric with respect to the plane of particle gyration. Consequently, radiation does not induce a recoil momentum in the parallel direction, so that $v_{\|}$ remains constant. See Appendix for a rigorous proof demonstrating that the parallel velocity remains constant. Also, we have 
${\partial\gamma}/{\partial v_{\bot}^{2}}=\gamma^{3}/2c^{2}.$
Thus, radiative cooling results in
\begin{equation}
\left.\frac{d v_{\bot}^{2}}{d t}\right|_{\rm rad}=-\frac{2P}{m\gamma^{3}}.
\label{dvperp-rad}
\end{equation}

Now, consider the magnetic moment equation. In the absence of radiative losses, it is given by Eq. (\ref{mu}). It can be re-written as
\begin{equation}
\frac{v_{\bot}^{2}}{B}\frac{d \gamma^{2}}{dt}
+\frac{\gamma^{2}}{B}\frac{d v_{\bot}^{2}}{dt}
+\gamma^{2} v_{\bot}^{2}\frac{dB^{-1}}{dt}=0.
\end{equation}
Furthermore
\begin{equation}
\frac{d\gamma^{2}}{dt}=
\frac{\partial\gamma^{2}}{\partial v_{\|}^{2}}\frac{d v_{\|}^{2}}{d t}
+\frac{\partial\gamma^{2}}{\partial v_{\bot}^{2}}\frac{d v_{\bot}^{2}}{d t}
=\frac{\gamma^{4}}{c^{2}}\left(\frac{d v_{\|}^{2}}{d t}+\frac{d v_{\bot}^{2}}{d t}\right).
\end{equation}
Now, we plug this into the previous equation and collect terms with ${d v_{\bot}^{2}}/{d t}$. We obtain the equation for the perpendicular velocity of a particle in an inhomogeneous field (e.g., in a magnetic mirror) set by the conservation of the magnetic moment, that is, in the absence of radiative losses:
\begin{eqnarray}
\left.\frac{d v_{\bot}^{2}}{dt}\right|_{\rm mag}
&=&-\left(\frac{v_{\bot}^{2}}{B}\frac{\gamma^{4}}{c^{2}}\frac{d v_{\|}^{2}}{d t}
+\gamma^{2} v_{\bot}^{2}\frac{dB^{-1}}{dt}\right)
\nonumber\\
&{ }&\times\left(\frac{v_{\bot}^{2}}{B}\frac{\gamma^{4}}{c^{2}}
+\frac{\gamma^{2}}{B}\right)^{-1}.
\label{dvperp-mu}
\end{eqnarray}
The denominator can be simplified further to be $\left(\gamma^{2}/B\right)
\left(1+v_{\bot}^{2}\gamma^{2}/{c^{2}}\right)=\gamma^{4}/\gamma_{\|}^{2}B$, where we introduced $\gamma_{\|}=1/\sqrt{1-v_{\|}^{2}/c^{2}}$. 

In the presence of radiative losses, we have
\begin{equation}
\frac{d v_{\bot}^{2}}{d t}
=\left.\frac{d v_{\bot}^{2}}{dt}\right|_{\rm mag}
+\left.\frac{d v_{\bot}^{2}}{d t}\right|_{\rm rad}.
\label{dvperp-E}
\end{equation}
Formally, this equation describes the evolution of $v_{\bot}$ we have been looking for. However, re-assembling the magnetic moment on the left hand side, one obtains a more physically transparent equation
\begin{equation}
\frac{d\left(m\gamma^{2} v_{\bot}^{2}/2B\right)}{dt}=\frac{d\mu_{\rm r}}{dt}
=-\frac{\gamma P}{\gamma_{\|}^{2} B}.
\label{mu'}
\end{equation}
This equation replaces Eq.~(\ref{mu}) in the case when the particle radiative (cyclotron or synchrotron) losses are not negligible.

\section{Evolution of the distribution of ``Larmor particles''}
\label{ensemble}

Frequently, we are interested in the behavior of an ensemble of particles rather than the trajectories of individual particles. A general particle distribution accounts for the positions and velocities (or momenta) of the particles, denoted as $f({\bf r},{\bf v})$. When the rapid gyro-motion of a particle is not of interest and is integrated out, we consider the motion of a Larmor circle, which is represented by the position and velocity of its center. This construct is referred to as the ``Larmor particle." 

The equations of motion for such a Larmor particle subject to energy loss are given by the system of equations, as follows. Eq.~\eqref{R} provides the position of the Larmor particle along the magnetic field line.  Eq.~\eqref{v} represents the equation of motion along the field under the influence of a mirror force. Eq.~\eqref{mu'} describes the evolution of the magnetic moment in the presence of energy loss, expressed through the power $P$. It is important to note that $P$ can encompass any form of energy loss, not just radiative loss in a magnetic field. The sole assumption made about $P$ is that this loss, when measured in the guiding center frame ($v_{\|}=0$), does not induce recoil along the magnetic field. In other words, the dissipated power is symmetrically distributed with respect to the plane perpendicular to the magnetic field. In our case, we consider the radiative energy loss via synchrotron/cyclotron emission, Eq.~\eqref{P}, which implicitly depends on $\mu_{\rm r}$ via the perpendicular velocity, $v_{\bot}$, as is given by Eq.~\eqref{mur}.

The distribution function of Larmor particles can be introduced straightforwardly as $f(\bar{\bf r},v_{\|},\mu_{\rm r})$. Its evolution in time can be obtained numerically. Given the initial distribution function $f(\bar{\bf r}_{0},v_{\|0},\mu_{\rm r0})$, one uses Eqs. \eqref{R}, \eqref{v}, and \eqref{mu'} to compute the subsequent values $\bar{\bf r},v_{\|},\mu_{\rm r}$ at time $t$, for each particle. Re-binning these values, we get the time-dependent distribution function $f(\bar{\bf r},v_{\|},\mu_{\rm r})$ at an arbitrary time. Note that Liouville’s theorem --- i.e., conservation of the distribution function along particle trajectories --- cannot be applied here (see Appendix). The phase-space volume is not conserved in the presence of cooling, because the radiation reaction force is dissipative.

\section{Motion in magnetic mirror with synchrotron losses}
\label{bottle}

\subsection{Model of a magnetic mirror}

Now, we investigate particle evolution numerically. We introduce dimensionless quantities: $x=\left|\bar{\bf r}\right|/R_{0}$, $\tau=t/t_{0}$, where $R_{0}$ is some reference distance in the neutron star magnetosphere and $t_{0}=R_{0}/c$ is the corresponding light crossing time. Furthermore, $\beta_{\|}=v_{\|}/c$, $\beta_{\bot}=v_{\bot}/c$, $\beta=v/c$, $\gamma_{\|}=\left(1-\beta_{\|}^{2}\right)^{-1/2}$, $\gamma_{\bot}=\left(1-\beta_{\bot}^{2}\right)^{-1/2}$, $\gamma=\left(1-\beta^{2}\right)^{-1/2}$. Next, we neglect the field line curvature, for simplicity. This is done for illustrative purpose only since. There is no problem in solving the ``drift-loss equations'' for an arbitrary field configuration. Since the gradient and curvature drifts are small, as we discussed in Section \ref{gc}, our results should be of general applicability to a neutron star magnetosphere. We can further assume that the NS magnetic field is described by a power-law in distance: $B=B_{NS}r^{-n}=B_{0}x^{-n}$, where $B_{NS}$ is the surface magnetic field at the pole of a NS, $r=\left|\bar{\bf r}\right|/R_{NS}$, $R_{NS}\sim10^{6}\textrm{ cm}$ is the neutron star size and $B_{0}=B_{NS}\left(R_{0}/R_{NS}\right)^{-n}$ is the magnetic field at distance $R_{0}$. For numerical convenience, we chose $R_{0}=100 R_{NS}=10^{8}\textrm{ cm}$. Thus, in these units $x=1$ corresponds to 100 neutron star radii. Hence, $\nabla B=-n\left(B_{0}/R_{0}\right)x^{-n-1}\hat{\bf b}$. With these definitions, the ``drift-loss equations'', Eqs.~(\ref{R}), (\ref{v}), (\ref{mu'}), read:
\begin{eqnarray}
\frac{d x}{d\tau}&=&\beta_{\|}, 
\label{R+}\\
\frac{d\left(\gamma \beta_{\|}\right)}{d\tau}&=&\frac{n}{2}\frac{\gamma \beta_{\bot}^{2}}{x},
\label{v+}\\
\frac{d\left(\gamma^{2} \beta_{\bot}^{2} x^{n}\right)}{d\tau}&=&-C_{R}\frac{\gamma^{3}\beta_{\bot}^{2}}{\gamma_{\|}^{2} x^{n}}.
\label{mu'+}
\end{eqnarray}
Here we've introduced the dimensionless `radiative constant' 
$$
C_{R}=\frac{\sigma_{T}}{2\pi }\frac{R_{0}B_{0}^{2}}{mc^{2}}
\simeq 1.28\times10^{5}\left(\frac{R_{0}}{10^{8}\textrm{ cm}}\right)^{-5}
\left(\frac{B_{NS}}{10^{14}\textrm{ G}}\right)^{2},
$$
where we assumed $n=3$ in the second equality, so that $B_{0}=10^{8}\textrm{ G}$. 
We also note that in the numerical results presented below, the particle momenta are dimensionless, as they are normalized by $mc^2$.

\begin{figure}
\includegraphics[scale = 0.9]{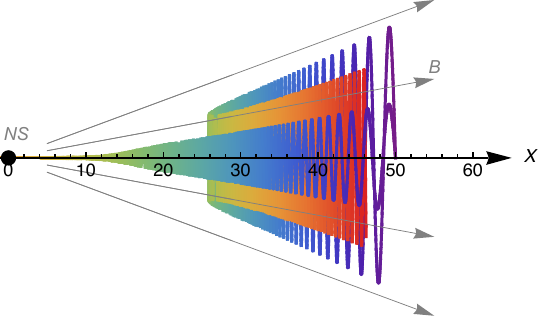}  
\caption{Schematic diagram showing the magnetic bottle. A neutron star (size is not to scale) is located at $x=0$ where the magnetic field is the strongest. Two particle trajectories starting at the same location, $x=50$, but with different pitch angles are shown for illustration. The vertical scale is arbitrary. Color gradient (blue-green-red) indicates time since the start, with the green color roughly corresponding to the reflection point in the absence of cooling. 
}
\label{cartoon}
\end{figure}

\begin{figure}
\includegraphics[scale = 0.92]{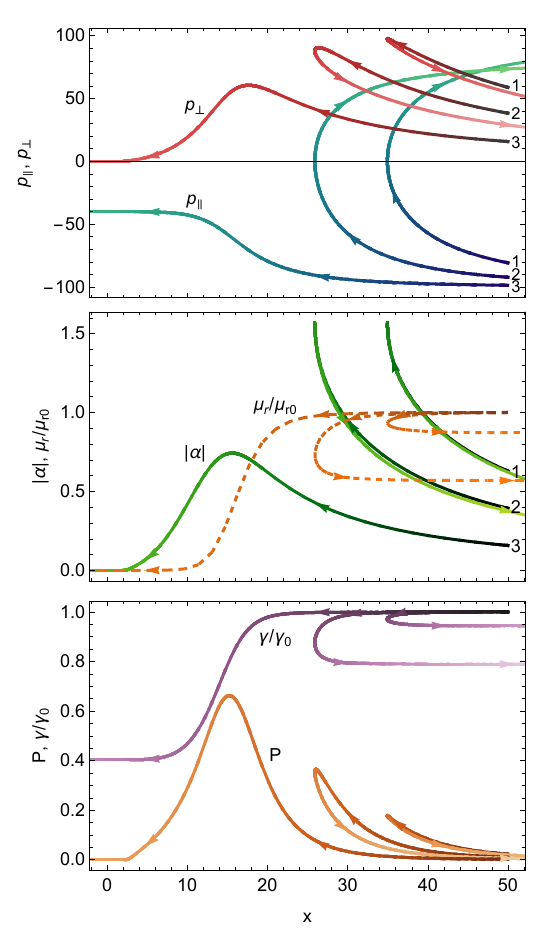}  
\caption{Evolution of  parameters of the particles propagating at three initial pitch angles: $\alpha_{0}=\pi/5,\ \pi/8,\ \pi/20$, labeled at the beginning ($\tau=0$) as 1, 2, and 3 respectively. The particles are injected at $x=50$ with the initial Lorentz factor $\gamma_{0}=100$. The evolution in time is shown by arrows and the color gradient: dark-to-light color change indicates early-to-late time evolution. Shown are: ({\em top panel}) the parallel components, $p_{\|}$, (blue-green colors) and perpendicular components, $p_{\bot}$, (red-pink colors) of particles' momenta; ({\em middle panel}) the absolute values of the pitch angle of the particles, $|\alpha|$, (green colors) and their relativistic magnetic moments normalized by their initial values, $\mu_{\rm r}/\mu_{\rm r0}$ (dashed lines, brown colors); ({\em bottom panel}) the particles' Lorentz factors normalized to their initial values, $\gamma/\gamma_{0}$, (pink colors) and the instantaneous emitted power, $P$ (orange colors, arbitrary units). The numerical labels for the curves representing $\mu_{\rm r}/\mu_{\rm r0},\ P$, and $\gamma/\gamma_{0}$, are omitted to avoid confusion and since attribution of these curves is straightforward.
}
\label{3angles}
\end{figure}

\begin{figure}
\includegraphics[scale = 0.9]{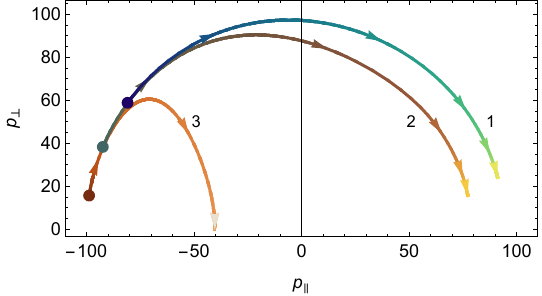}  
\caption{Evolution of the momenta of the same particles as in Fig. \ref{3angles} in the $p_{\|}$--$p_{\bot}$ momentum space. The initial momenta are shown with big points and roughly located at $p_{\bot}\sim60, \ 40,\ 20$ for particles 1, 2, and 3, respectively. Time evolution is shown with arrows and color gradients (from dark to light). 
}
\label{3pp}
\end{figure}

This configuration is illustrated in Figure \ref{cartoon}. The neutron star is represented by a large dot (not to scale), positioned at $x=0$. Two particle orbits are depicted, with their absolute sizes and gyro-periods are arbitrary and mostly serve as indicators of relative changes. . These orbits illustrate the salient features of the particles' trajectories. Both particles start their motion at $x=50$ (that is, at the radius $5000 R_{NS}$) and move inward. Their initial Lorentz factor is $\gamma_{0}=100$ and initial pitch-angles are $\alpha=\pi/8,\ \pi/11$. Time is indicated by the color: early time corresponds to violet and blue hues, green tones approximately represent the time of reflection (in the absence of cooling), and yellow and red hues correspond to late times.

Two types of trajectories are observed: `{\em trapped}' trajectories and `{\em precipitating}' trajectories. Trapped trajectories are the standard orbits of trapped particles that are now subject to radiative energy loss. In contrast, precipitating particles, which would have also been trapped in the absence of energy loss, are no longer reflected back. This is so because the reflection point of the precipitating particle is located within the region of the very strong magnetic field. Consequently, the particle loses its perpendicular energy too rapidly to be reflected. In other words, its magnetic moment (and hence the mirror force) diminishes to zero too fast, before the particle is reflected. Due to the retention of some parallel momentum, the particle continues its forward motion and ultimately collides with the neutron star. In the subsequent figures (Fig.~\ref{3angles}, \ref{3pp}), these two trajectories are labeled as `2' and `3', respectively.

\subsection{Single particle dynamics}

Figure \ref{3angles} illustrates the temporal evolution of various parameters for three particles. As before, the particles begin their motion at $x=50$ and move inward, hence their parallel momenta are initially negative. Time is represented by arrows and color gradients, transitioning from dark tones indicative of early times to lighter tones representing late times. The particle trajectories are labeled by numbers at the beginning, at $\tau=0$. The initial energy of the three particle is the same $\gamma_{0}=100$ but their pitch angles differ. Particles 1 and 2 are the {\em trapped} particles, particle 1 is only weakly cooled, whereas particle 2 experiences stronger cooling loss.  Particle 3 is the {\em precipitating} particle. 

The top panel illustrates the evolution of the parallel (blue-toned) and perpendicular (red-toned) momenta. The parallel momenta of particles 1 and 2 undergo a sign change, becoming positive, which signifies magnetic reflection. These reflections occur approximately at $x=35$ and $x=25$, respectively. Conversely, the parallel momentum of particle 3 remains negative, while its perpendicular momentum vanishes. Consequently, the particle continues its motion towards the origin and ultimately collides with the neutron star's surface. 

The middle panel depicts the evolution of the magnitude of the pitch angle $|\alpha|=\left| \arctan\left(p_{\bot}/p_{\|}\right)\right|$ (green hues) and the relativistic magnetic moment (dashed curves, brown hues), as given by Eq. \eqref{mur}, normalized to its initial value $\mu_{\rm r0}=\mu_{\rm r}(\tau=0)$. Notably, the pitch angle attains the value of $\pi/2$ at the reflection points of particles 1 and 2. In contrast, the precipitating particle's $|\alpha|$ never reaches this value. The magnetic moment evolution is also straightforward. It serves as an adiabatic invariant in the absence of energy loss. However, with energy loss, $\mu_{\rm r}$ remains constant when radiative loss is minimal, but begins to diminish significantly in the strong field region, in accordance with Eq. \eqref{mu}. For particle 3, $\mu_{\rm r}$ and $|\alpha|$ completely vanish at approximately $x=3$. 

The bottom panel illustrates the evolution of the particle's Lorentz factor, normalized to its initial value, $\gamma/\gamma_0$, and the instantaneous emission power, $P$ (in arbitrary units), as described by Eq. \eqref{P}. Notably, the trapped particles (1 and 2) experience a significant energy loss near the reflection points. Conversely, particle 3 undergoes a severe energy loss prior to reflection. Therefore, it continues its motion into the stronger field region, resulting in even more pronounced cooling. Overall, this runaway process culminates in catastrophic energy loss within a finite time frame. Consequently, among the two types of particles, the precipitating ones exhibit the highest efficiency in terms of emitted radiation power, as they dissipate all their energy associated with gyro-motion within a short duration. 

It is also instructive to present the evolution of particles in momentum space. Figure \ref{3pp} illustrates the same particles (1, 2, and 3) in the $p_{\bot}$ versus $p_{\|}$ diagram. It is evident that the particles originate at three distinct points on a circle, $\gamma^{2}=p_{\|}^{2}+p_{\bot}^{2}$, at three different pitch angles. They subsequently traverse the circle until their energy loss becomes substantial. As before, the time evolution is depicted by arrows, with lighter tones signifying later times. Three distinct color tones represent different particles.

\begin{figure}
\includegraphics[scale = 0.9]{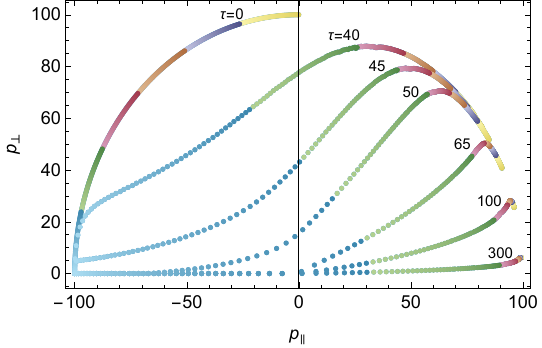}  
\caption{Time evolution of an initially monoenergetic isotropic particle distribution plotted in momentum space. Initially, the particles have the Lorenz factor of $\gamma_{0}=100$ and are equally spaced in pitch-angle. The curves are labeled by time: $\tau=0,\ 40,\ 45,\ 50,\ 65,\ 100,\ 300$. Each point represents a particle. The particles are colored by groups (of equal total number in each group) to ease understanding the graph. Note that large-pitch-angle particles (yellow, violet, orange) are reflected backwards in the low-field region and do not lose much energy (they remain on the circle $p_{\|}^{2}+p_{\bot}^{2}=\gamma^{2}_{0}$), whereas those with small pitch-angle penetrate into the high-field region and lose their transverse momentum quickly. 
}
\label{circ}
\end{figure}

\begin{figure}
\includegraphics[scale = 0.9]{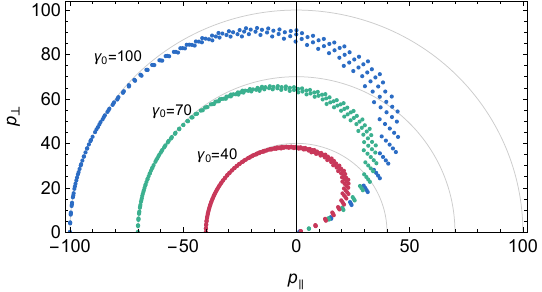}  
\caption{Steady-state $p_{\|}$--$p_{\bot}$-momenta distributions (each point represents a particle), being injected with isotropic pitch angles at $x=50$ and with $\gamma_{0}=100,\ 70,\ 40$, and measured in the interval $25\le x\le 28$. Thin semi-circles are the lines of constant $\gamma$, to guide the eye. 
}
\label{pdf3}
\end{figure}

\begin{figure*}
\includegraphics[scale = 1.15]{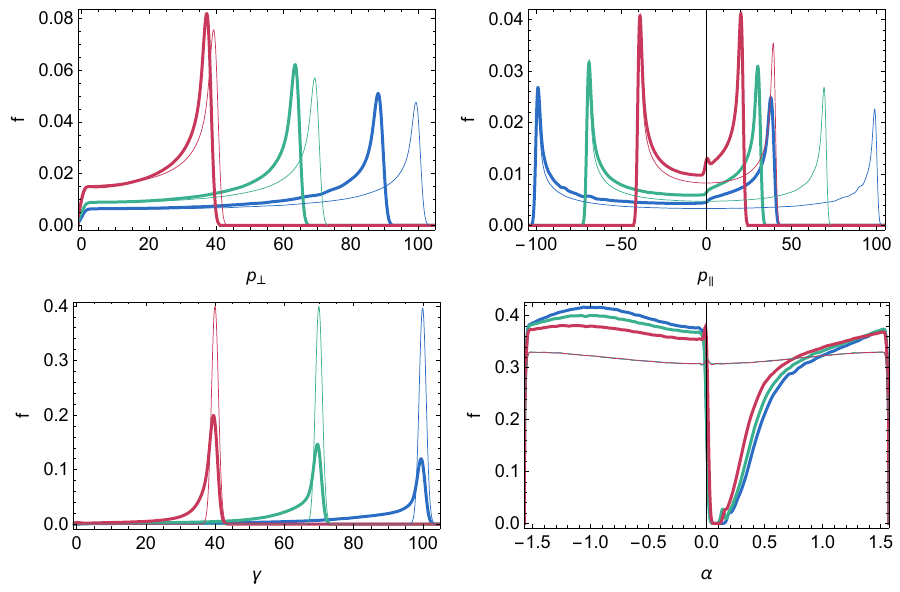}  
\caption{Various representations of the same distributions, $f(p_{\|},p_{\bot})$, as in Figure \ref{pdf3}. The top row shows $f(p_{\bot})$ and $f(p_{\|})$, i.e., the distributions integrated over the parallel and perpendicular momenta, respectively. The bottom row shows the energy and pitch-angle distributions, $f(\gamma)$ and $f(\alpha)$. Thick curves represent the case with cooling and thin curves depict the no-cooling case. 
}
\label{pdf34}
\end{figure*}

\begin{figure}
\includegraphics[scale = 1.05]{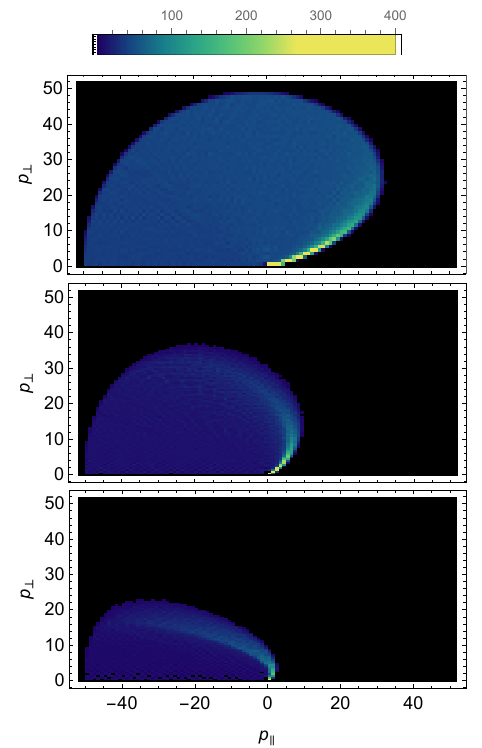}  
\caption{Steady-state particle distribution function of a top-hat distribution being injected at $x=50$ with $\gamma_{\rm max}=50$, and measured at $x=30,\ 15,\ 10$. Color coding is arbitrary, but brighter colors indicate a higher particle density.
}
\label{pdf}
\end{figure}

\begin{figure}
\includegraphics[scale = 1.0]{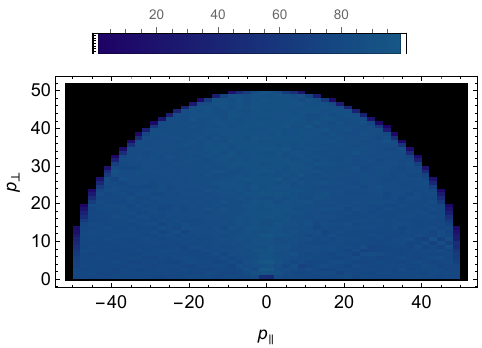}  
\caption{Steady-state particle distribution function as in Fig. \ref{pdf} (and measured at $x=30$) but in the absence of cooling. The loss-cone located at positive $p_{\|}$ is too small to be seen. }
\label{pdf1no}
\end{figure}

\subsection{Particle distribution function dynamics}

As a subsequent illustration, we demonstrate the evolution of an isotropic monoenergetic distribution function. Figure \ref{circ} depicts 300 particles at various times. The particles are divided into six groups, each comprising 50 particles, and colored appropriately to facilitate the observation of their evolution. Initially, the particles possess a Lorentz factor of $\gamma_{0}=100$ and are located at $x=50$. The particles are uniformly distributed over the pitch angle along a quarter-circle, with $p_{\bot}>0$ and $p_{\|}<0$ (indicating motion to the left, towards stronger magnetic fields). Notably, particles with large pitch angles (colored in yellow, violet, and orange) are reflected in a low-field region and experience minimal energy loss (at later times, they predominantly overlap on the graph). Consequently, they remain confined to the circle defined by $p_{\|}^{2}+p_{\bot}^{2}=\gamma^{2}_{0}$. This behavior is characteristic of magnetically trapped particles. The group of particles colored in red starts to deviate from this circle, suggesting mild energy loss during reflection. The green group of particles (with sufficiently small initial pitch angles) undergoes substantial energy loss. Despite this, they remain trapped because all possess positive $p_{\|}$ at late times. Finally, the majority of the particles in the blue group experience catastrophic radiative losses and precipitate, as $p_{\|}$ remains negative throughout the entire evolution. Only a small fraction of these particles is reflected, but they lose almost all of their kinetic energy. 

To this point, we have been considering an initial value problem. In this context, we have a specific distribution created at time $\tau=0$ and seek to understand its evolution over time. Now, we transition to a steady-state problem. This involves assuming a continuously operating source of particles at a particular location and investigating the type of particle distribution that forms at various locations. We note that it suffices to assume that the source's operation must occur on a time scale longer than the bounce time. Furthermore, this scenario is equivalent to the initial value problem discussed previously, provided that we are interested in the time-integrated distribution function observed at a given location.

Figure \ref{pdf3} illustrates three steady-state distributions. The particle source is situated at $x=50$ and continuously injects particles with three distinct energies: $\gamma_{0}=100,\ 70$, and $40$. The injected distributions are isotropic over the pitch angle in the range $0\le\alpha\le\pi/4$ subdivided into 100 intervals, yielding the numerical resolution in the pitch-angle of $\Delta\alpha=\pi/400$. The resultant steady-state particle distribution is observed within the interval $25\le x\le 28$. The figure presents the complete distributions as a function of momenta, $f(p_{\|},\, p_{\bot})$. Figure \ref{pdf34} shows various representations of  $f(p_{\|},p_{\bot})$. The top row presents $f(p_{\bot})$ and $f(p_{\|})$, that is, the distributions integrated over the parallel momenta and perpendicular momenta, respectively. The bottom row presents the energy and pitch-angle distributions, $f(\gamma)$ and $f(\alpha)$. Thick curves represent the case with radiative cooling and thin curves correspond to the case without cooling. A distinct {\em loss-cone-like} distribution emerges, differing from the standard loss-cone distribution without cooling. Notably, the cone opening angle in this case is (i) significantly larger, (ii) energy-dependent, and (iii) controlled by cooling. For comparison, without cooling, the loss cone opening angle would be $\sin\alpha(x)=B(x)/B(0.01)$, where $B(0.01)$ represents the field strength at the neutron star surface. In our case, with $x=25$ and $B\propto x^{-3}$, this angle would be minuscule, $\alpha(25)\sim 6\times 10^{-11}$, in contrast to the observed value of $\alpha\sim\pi/4-\pi/6$, depending on energy. 

Whereas Figure \ref{pdf3} effectively illustrates the spatial distribution of particles in momentum space, it provides limited information regarding their momentum-space density. To investigate the latter, we now set the source to generate a ``top-hat" particle distribution, defined as $f(p_{\|},\, p_{\bot})=const.$ for $\gamma\le\gamma_{\rm max}$ and zero otherwise. We set the maximum Lorentz factor to be $\gamma_{\rm max}=50$. The resulting steady-state distribution is depicted in Figure \ref{pdf} at three distinct locations: $30\le x\le32$ (top panel), $15\le x\le17$ (middle panel), and $10\le x\le12$ (bottom panel). The color-coding represents the momentum-space particle density (arbitrary units) with a linear scale, wherein brighter tones signify higher density. For comparison, Figure \ref{pdf1no} presents the distribution function at $x=30$ in the absence of cooling. It is evident that the distribution is uniform and isotropic at $\gamma\le\gamma_{\rm max}=50$. The loss-cone $\alpha\sim10^{-11}$, located at $p_{\|}>0$, is too small to be seen. 

We note two points in this context. Firstly, the distribution bears a resemblance to the loss-cone distribution, characterized by a cone-shaped region surrounding the positive $p_{\|}$ axis that is devoid of particles. However, the cone opening angle $\alpha(\gamma)$ is energy dependent, in contrast to the non-radiative case. Secondly, the region of highest particle density is situated precisely at the periphery of the cone, surpassing the mean density by a substantial factor of 5 to 10 in our case, depending upon the specific location. Consequently, this distribution can be likened to a funnel positioned along the $p_{\|}$ direction, with the latter serving as the axis of symmetry. Therefore, we can refer to it as the ``{\em funnel distribution}" or the ``cooled-loss-cone'' distribution.

\subsection{Analytical estimates}

The most intriguing aspect lies in the interplay between the mirroring force and radiative cooling. In essence, if the cooling time is shorter than the particle's residence time in the mirror, i.e., $t_{\rm cool} \ll t_{\rm mirr}$, then such a particle will rapidly lose its energy. This phenomenon is referred to as the ``{\em fast-cooling regime}" and is anticipated to occur deep within the neutron star magnetosphere, where the magnetic field strength is exceptionally high. Conversely, the case where $t_{\rm cool} \gg t_{\rm mirr}$ corresponds to the ``{\em slow-cooling regime}," which is expected to exist at a considerable distance from the neutron star's surface. We now estimate the critical ``cooling" radius, $R_{c}$, at which these two time scales become comparable, i.e., $t_{\rm cool} \simeq t_{\rm mirr}$.

A characteristic time a trapped particle requires to transfer most of its parallel energy into its perpendicular motion within the strong magnetic field of a magnetic mirror and than back into the parallel motion in the opposite direction is readily estimated from Eq. \eqref{v}:
\begin{equation}
\frac{2(m\gamma v_{\|})}{t_{\rm mirr}}\simeq \frac{\mu_{\rm r}}{\gamma} |\nabla B|,
\end{equation}
Consequently, the mirroring time is 
\begin{equation}
t_{\rm mirr}\simeq \frac{4L}{c}\frac{\beta_{\|}}{\beta_{\bot}^{2}},
\end{equation}
which is about four times the light crossing time. Here we introduced the characteristic scale of the $B$-field inhomogeneity, $L\equiv B/|\nabla B|=R/n$, for a magnetic field $B\propto R^{-n}$. 

The cooling time follows from Eqs.~\eqref{E}, \eqref{P}:
\begin{equation}
t_{\rm cool}\simeq \frac{4\pi m c}{\sigma_{T}}\left(\beta_{\bot}^{2}\gamma B^{2}\right)^{-1}
\approx (5.16\times 10^{8}\textrm{ s})\left(\beta_{\bot}^{2}\gamma B^{2}\right)^{-1},
\end{equation}
where $B$ is in gauss. The condition for strong cooling $t_{\rm cool} \ll t_{\rm mirr}$ takes the form
\begin{equation}
\beta_{\|}\gamma B^{2}L\gg \pi m c^{2}/\sigma_{T}\approx 3.87 \times 10^{18}\textrm{ (cgs units)}.
\end{equation}

At the critical ``cooling" radius, $t_{\rm cool} \simeq t_{\rm mirr}$. Using that $B=B_{NS}(R/R_{NS})^{-3}$ and $L=R/3$, we obtain the cooling radius
\begin{equation}
\frac{R_{c}}{R_{NS}}=\left(\frac{\sigma_{T}\beta_{\|}\gamma B_{NS}^{2}R_{NS}}{3\pi m c}\right)^{1/5}\simeq 973\, \beta_{\|}^{1/5}\gamma^{1/5} B_{NS,14}^{2/5},
\label{Rc}
\end{equation}
where $B_{NS,14}=B_{NS}/(10^{14}\textrm{ gauss})$ and $R_{NS}\simeq10$~km. It is noteworthy that the cooling radius exhibits a relatively modest sensitivity to the particles' Lorentz factor and falls within a range of approximately a few hundred to a thousand neutron star radii, under the typical conditions around pulsars and magnetars. Notably, this distance corresponds to $x\sim 10$ in our numerical analysis. This value remarkably aligns with the location of the region in Figure \ref{3angles} where cooling is most pronounced, specifically around $x\sim 15$.

The strong cooling regime occurs at $R<R_{c}$, that is in the inner magnetosphere, generally inside the light cylinder,
\begin{equation}
R_{LC}=c/\Omega_{\rm rot}\simeq (4800\, R_{NS})P_{\rm rot},
\end{equation}
where $\Omega_{\rm rot}$ and $P_{\rm rot}$ are the angular frequency and the period (in seconds) of rotation of a neutron star. Within the cooling region ($R<R_{c}$), the particles experience rapid energy loss, resulting in a minimal confinement by the magnetic mirroring force. Consequently, these particles precipitate towards the surface. In contrast, outside this region ($R>R_{c}$), the particles are weakly cooled, so they can maintain their distribution for many light-crossing times. In particular, these trapped particles may experience many bounces off the magnetic mirrors in the polar regions. Thus, a quasi-steady-state, gradually cooling  distribution ({\em a l\`a} van Allen belts) with ``cooled loss cones'' should form.

\begin{figure}
\includegraphics[scale = 0.88]{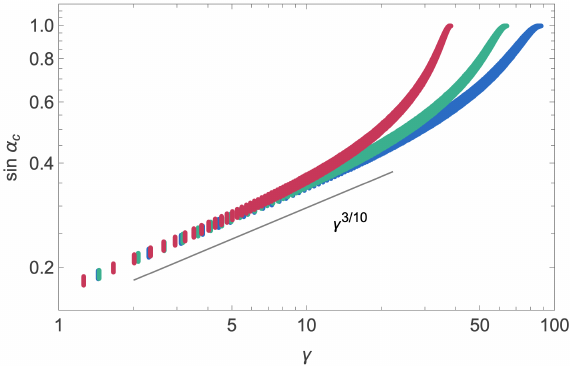}  
\caption{Logarithmic plot of the sine of the pitch angle at $x=25$ versus the particle's Lorentz factor, for the same distributions as in Figure \ref{pdf3}. The thin line indicates the scaling given by Eq.~\eqref{ac} and valid at small angles.
}
\label{cone}
\end{figure}

The size of the loss cones is evaluated as follows. For trapped particles far away from $R_{c}$, radiative losses are negligible, resulting in the conservation of the magnetic moment, $\mu_{\rm r}=mc^{2}\gamma^{2}\beta_{\bot}^{2}/2B=const$. Since the pitch angle is $\sin\alpha=p_{\bot}/p=(\gamma\beta_{\bot})/\gamma=\beta_{\bot}$, we have the following well-known expression: 
\begin{equation}
\sin^{2}\alpha(R)=B(R)/B(R_{r}),
\label{sin2a}
\end{equation}
where $R_{r}$ is the reflection point at which $\sin\alpha(R_{r})=1$, provided that $R_{r}>R_{c}$ (otherwise, the particle precipitates, rather than being reflected back). It is important to note that $B(R)\le B(R_{r})\ll B(R_{c})$ and hence $\alpha(R)\ll1$, for the slow-cooling approximation to be valid. On the other hand, the particles originating within the loss cone are those that pass through $R_{c}$ and subsequently precipitate. Consequently, the boundary of the cooled loss cone, $\alpha_{c}$, is determined by the condition $R_{r}\simeq R_{c}$. By substituting Eq.~\eqref{Rc} into Eq.~\eqref{sin2a}, we obtain the angle of the cooled loss cone at a specific location $R$ as
\begin{equation}
\sin\alpha_{c}(R)=\left(\frac{R_{c}}{R}\right)^{3/2}
\simeq B_{NS,14}^{3/5}\gamma^{3/10}\left(\frac{R}{10^{3}\,R_{NS}}\right)^{-3/2}.
\label{ac}
\end{equation}
This expression is accurate provided that $\alpha_{c} \ll 1$, so we utilized that $\beta_{\|}=\cos\alpha_{c}\approx 1$. Figure \ref{cone} illustrates the relationship between $\sin \alpha$ and the particle's energy for the same three distributions as depicted in Figure \ref{pdf3}. It is evident that, at low pitch angles, all the distributions closely follow the analytical result given by Eq. \eqref{ac}. Notably, the cooled loss cone angle exhibits energy dependence and scales as 
\begin{equation}
\alpha_{c}\propto \gamma^{3/10}.
\end{equation}

It is also interesting to note the strength of the magnetic field in which the synchrotron power is the largest,
\begin{equation}
B(R_{c})\simeq (10^{5}\textrm{ G})\, B_{NS,14}^{-1/5}\gamma^{-3/5},
\end{equation}
is very insensitive to the neutron star surface field (here we assumed $\beta_{\|}\simeq1$). The total energy in this field is
\begin{equation}
E\sim (B^{2}/8\pi)(4\pi/3)R_{c}^{3}\sim (10^{36}\textrm{ erg})\,B_{NS,14}^{4/5}\gamma^{-3/5}.
\end{equation}
This expression represents a reasonable upper limit on the energy that can be converted into photons within the light-crossing time, $t_c\sim R_c/c\sim 10-30$~ms, assuming that the primary source of free energy is the local magnetic field. If the produced funnel particle distribution can produce a maser emission, its characteristic frequency is expected to be around the relativistic cyclotron frequency
\begin{equation}
\nu_{ce}=\frac{e B}{\gamma mc}\simeq (180\textrm{ MHz})\,B_{NS,14}^{-1/5}\gamma_2^{-8/5},
\end{equation}
where $\gamma_2=\gamma/10^2$. Thus, the emission is expected to be in the radio band.

\section{Main results}
\label{concl}


First, we derived the evolution equations that describe the motion of a relativistic particle within a magnetic bottle, assuming that radiative energy losses cannot be neglected, Eqs.~\eqref{R}, \eqref{v}, \eqref{mu'}. These equations are derived using the guiding center approximation and neglecting particle drifts and should be useful for modeling particle dynamics in pulsar and magnetar magnetospheres subject to synchrotron cooling. They should also be useful for understanding the plasma populations in these magnetospheres, including the trapped populations akin to van Allen belts.  

Second, we investigated the typical dynamics and identified the presence of two distinct types of trajectories. Trajectories with large initial pitch angles form a relatively stable trapped population that gradually loses its energy, primarily near their magnetic reflection points. In contrast, trajectories with small pitch angles result in catastrophic energy loss (i.e., within a finite time) by the particles, preventing their reflection and ultimately leading to their precipitation onto the star's surface.

Third, in the presence of a source of energetic particles that operates on a time scale exceeding a single magnetic reflection, a steady-state distribution emerges. This distribution bears a resemblance to a loss-cone distribution, with the energy-dependent opening angle influenced by radiative cooling. Consequently, it can be referred to as the ``cooled-loss-cone" distribution. However, the momentum space density exhibits significant non-uniformity. The region of highest particle density is situated at the edge of the loss cone, surpassing the mean density by a substantial factor of at least a few or more. Consequently, the distribution can also be termed as the ``funnel" distribution. The size of the loss cone is energy-dependent and scales as $\alpha_{c} \propto \gamma^{3/10}$. It is also demonstrated that synchrotron losses are the strongest in a well-localized region of the magnetosphere, approximately between a few hundred and a thousand star radii under typical conditions of pulsars and magnetars. Specifically, $R_{c}\simeq (10^{3}R_{NS})\gamma^{1/5} B_{NS,14}^{2/5}$. The synchrotron emission emanating from this region can be observed as non-polar emission emitted from the outer magnetosphere. 

It is widely recognized that the loss-cone distribution is unstable and capable of generating maser radiation, \citep{GZh70, Zh97book}. The substantially elevated particle density observed in the funnel distribution (in comparison to the loss-cone distribution) can only enhance the efficiency of this maser. It is plausible that such a maser is responsible for the coherent outer magnetospheric radiation and possibly weak fast radio bursts, such as those from the galactic magnetar. Indeed, the estimated upper limit on radiated energy and power from a millisecond-duration burst are $E\sim 10^{36}$~erg and $P\sim 10^{39}$~erg/s, respectively. The radiation is expected to be in the $\nu\sim0.2-7\textrm{ GHz}$ range for particles' Lorentz factors around $\gamma \sim10-100$. The presence of two magnetic mirrors within the magnetosphere can account for the occurrence of pairs (or, occasionally, groups of multiple bursts) separated by tens of milliseconds. This time scale can be interpreted as the duration of light and particle travel time between the regions where the cooling is strong and funnel distribution is formed (presumably, from hundreds to many thousands of stellar radii). Indeed, consider a scenario where a localized group (bunch) of energetic particles is produced (e.g., during a reconnection event) within the outer magnetosphere on the closed field lines. With time, these particles spread along the field lines towards both poles and enter the two strong field regions where they can produce two distinct emission episodes. However, depending on the initial parameters, not all the particles' energy may be radiated away in the first passage. In such cases, multiple bounces of these trapped particles off both mirrors can occur, leading to multiple episodes of enhanced emission.  Further numerical simulations are required to test this scenario.

In this paper, we examined a simplified model of the magnetosphere featuring a single magnetic mirror. In reality, the influence of both mirrors on the long-term evolution of particles, in conjunction with a more precise field geometry, should be considered. While such a study is intriguing and significant, it goes beyond the scope of the current paper.


\section*{Acknowledgements}

MM acknowledges the support by grant NSF PHY-2409249. AS and AP are supported by grants from NSF (PHY-2206607) and the Simons Foundation (MPSCMPS-00001470).
This research was supported in part by grant NSF PHY-2309135 to the Kavli Institute for Theoretical Physics (KITP).

\appendix

It is important to dive deeper into the constancy of parallel velocity. The radiation reaction force exerted on a relativistic particle acts in the direction opposite to its momentum. Therefore, it reduces both the perpendicular and parallel components of the particle's momentum. Consequently, the pitch angle of the particle, defined by the ratio of these components, remains constant during synchrotron cooling. However, despite $p_{\|}$ not being conserved, $v_{\|}$ is conserved, as elucidated in the text. Here, we rigorously prove this fact.

The synchrotron drag force (i.e., the radiation reaction force)  is typically written in the so-called Landau-Lifshitz form (see \citealp{LLvol2}, p. 213). Given the 4-velocity of the particle, $\gamma{\boldsymbol \beta}$, and the electric, and magnetic fields at the location of the particle, ${\bf E}$, and ${\bf B}$, the force can be written as:
\begin{equation}
\mathbf{F}_{RR}=\frac{\sigma_{T}c}{4\pi}
\Bigl[\bigl[\left({\bf E}+ {\boldsymbol\beta}\times{\bf B}\right)\times{\bf B}+ \left({\boldsymbol\beta}\cdot{\bf E}\right){\bf E}\bigr]-
\gamma^{2}{\boldsymbol\beta}\bigl[\left({\bf E}+ {\boldsymbol\beta}\times{\bf B}\right)^{2}-
\left({\boldsymbol\beta}\cdot{\bf E}\right)^{2}\bigr]\Bigr],
\end{equation}
where $\sigma_{T}\equiv(8\pi/3)r_{e}^{2}\simeq 6.65\times10^{-25}\textrm{ cm}^{2}$ is the Thomson cross-section. In our case, ${\bf E}=0$, so the expression for the force reduces to
\begin{equation}
\mathbf{F}_{RR}=-\frac{\sigma_{T}c}{4\pi}
\Bigl[({\bf B})^{2}{\boldsymbol\beta}-({\bf B}\cdot{\boldsymbol\beta}){\bf B}+\gamma^{2}({\boldsymbol\beta}\times{\bf B})^{2}{\boldsymbol\beta}
\Bigr].
\end{equation}
Its component parallel to the magnetic field is 
\begin{equation}
{F}_{RR, \|}\equiv\frac{\mathbf{F}_{RR}\cdot{\bf B}}{B}
=-\frac{\sigma_{T}c}{4\pi B}\ \gamma^{2}\left({\boldsymbol\beta}\times{\bf B}\right)^{2} \left({\boldsymbol\beta}\cdot{\bf B}\right)
=-\frac{\sigma_{T}c}{4\pi}\ \gamma^{2}\beta_{\bot}^{2}\beta_{\|}B^{2}.
\end{equation}
On the other hand, 
\begin{equation}
mc^{2}\frac{d(\gamma\beta_{\|})}{dt}
=mc^{2}\left(\dot\gamma\beta_{\|}+\gamma\dot\beta_{\|}\right)
={F}_{RR, \|}
=\beta_{\|}\dot {E}_{\rm rad} 
\end{equation}
because synchrotron power is $\dot {E}_{\rm rad}=-\dot\gamma mc^{2}=({\sigma_{T}c}/{4\pi})\gamma^{2}\beta_{\bot}^{2}B^{2}$. Therefore, the term $\gamma\dot\beta_{\|}=0$ identically, and hence $\beta_{\|}=const.$

Note that the dissipative nature of radiation reaction precludes the direct application of Liouville’s theorem to the evolution of the distribution function. Nevertheless, the distribution function still satisfies a continuity equation in 6D phase space:

\begin{equation}
\frac{\partial f(\mathbf{r}, \mathbf{p}, t)}{\partial t}
+ \frac{\partial}{\partial \mathbf{r}} \cdot (\mathbf{v} f)
+ \frac{\partial}{\partial \mathbf{p}} \cdot \left[ (\mathbf{F}_L + \mathbf{F}_{RR}) f \right]
= 0,
\end{equation}

where $\mathbf{F}_L$ and $\mathbf{F}_{RR}$ denote the Lorentz and radiation reaction forces, respectively. This leads to

\begin{equation}
\frac{{\rm d} f}{{\rm d} t}
= - f \cdot{\rm div}_{\mathbf{p}}( \mathbf{F}_{RR})\neq0,
\end{equation}

where ${\rm d}/{\rm d}t$ is the total derivative along the characteristics.

\bibliographystyle{aasjournalv7} 
\bibliography{lp-cool}    



\end{document}